
\magnification=\magstep1

\hsize=6truein
\vsize=8.5truein
\raggedbottom
\nopagenumbers
\tolerance=10000

\font\bfs=cmbx7

\font\bigbf=cmbx10 scaled \magstep1
\font\bigbold=cmbx10 scaled \magstep2

\def\skip{\vskip.15truein}

\def\today{\ifcase\month\or January\or February\or March\or April\or May\or
June\or July\or August\or September\or October\or November\or December\fi
   \space\number\day, \number\year}

\def\pmb#1{\setbox0=\hbox{#1}%
     \kern-.025em\copy0\kern-\wd0
     \kern.05em\copy0\kern-\wd0
     \kern-.025em\raise.0433em\box0 }

\nopagenumbers

\headline={\ifnum\pageno=1 \hss{\bigbf } \hss
  \else{\bfs Anderson, Random-Walks \hfil page \folio}\fi}
 \voffset=2\baselineskip

\hbox{ }
\vskip-20pt
\centerline {\bigbold Biased Random-Walk Learning:}
\smallskip
\centerline {\bigbold A Neurobiological Correlate to Trial-and-Error}
\skip
\centerline {Russell W. Anderson}
\centerline {Theoretical Biology and Biophysics}
\centerline {and The Center for Nonlinear Studies}
\centerline {Los Alamos National Laboratory}
\centerline {Los Alamos, NM  87545}
\centerline {email: rwa@temin.lanl.gov}
\centerline {(505) 667-9455}
\vskip.2truein
\baselineskip=15pt

\centerline {For copies of the figures,}
\centerline {please write or FAX R. Anderson}

\centerline {IN PRESS: {\it Progress in Neural Networks}}
\skip
\leftline{\bigbf ABSTRACT}
\skip

Neural network models offer a theoretical testbed for
the study of learning at the cellular level.
The only experimentally verified learning rule,
Hebb's rule, is extremely limited in its ability
to train networks to perform complex tasks.
An identified cellular mechanism responsible for
Hebbian-type long-term potentiation, the NMDA receptor,
is highly versatile.
Its function and efficacy are modulated by a wide variety
of compounds and conditions and are likely to be
directed by non-local phenomena.
Furthermore, it has been demonstrated that NMDA receptors
are not essential for some types of learning.
We have shown that another neural network learning
rule, the chemotaxis algorithm, is theoretically much more powerful
than Hebb's rule
and is consistent with experimental data.
A biased random-walk in synaptic weight space is
{\it a learning rule immanent in nervous activity} and
may account for some types
of learning -- notably the acquisition of skilled movement.

\skip
KEY WORDS: biological neural networks, random walk, chemotaxis,
stochastic optimization, biological plausibility.

\skip
\leftline{\bigbf INTRODUCTION}
\skip

\skip
In their landmark paper,
``A Logical Calculus of the Ideas Immanent in Nervous Activity",
McCulloch and Pitts [1943] demonstrated how a network of
extremely simplified (``all-or-nothing") neurons could compute any
Boolean function.
Mathematical analyses of modern
neural network models have since revealed them to be
potentially {\it universal} computing devices
[Seigelman and Sontag 1991].

\skip
Neural network modeling has not only
been helpful in understanding the collective
behavior of existing networks,
but also provides
a theoretical framework with which one can
experiment with models of learning.
Rosenblatt [1958] demonstrated that these networks,
when endowed with modifiable connections (``perceptrons"), could be
``trained" to classify patterns (see also Arbib [1964; 1987]).
Thus, Rosenblatt had developed a theoretical testbed
for the study of learning
(formerly the near-exclusive domain of psychology)
at the {\it cellular level}.

\skip
Theoretical neural network
studies (mathematical analyses and
empirical computer simulations)
are useful for exploring the capabilities and
limitations of a proposed learning rule.
The only experimentally verified learning
rule, Hebb's rule, has profound limitations
in this respect.
Engineering optimization algorithms
(such as back-propagation or genetic algorithms)
are capable of training neural networks
to perform much more sophisticated tasks,
but are biologically implausible
[Crick 1989a,b; Mel 1990; Anderson 1991].

\skip
Long underestimated by both the experimental and
theoretical neural network communities is
perhaps the most intuitive mode of learning --
trial-and-error.
We have shown [Bremermann and Anderson 1989,1991]
that the mathematical analog to trial-and-error,
a Gaussian biased random-walk in synaptic weight
space, is capable of training neural networks
to perform the same complex, nonlinear mappings
as backpropagation.

\skip
In this paper,
we review theoretical and empirical neural network
studies of random-walk learning which demonstrate
the effectiveness of this learning rule.
We argue the biological plausibility
of a trial-and-error learning rule,
though a discussion of existing
neurobiological data and identified molecular mechanisms.
Finally, we identify the directions of experimental
research most likely to
identify its necessary elements.

\skip
\leftline{\bigbf 2. HEBB'S RULE}

\skip
In 1949, Hebb proposed a neuronal learning rule which could
integrate associative memories into neural networks [Hebb 1949].
Hebb postulated that when one neuron repeatedly excites another,
the synaptic knobs are strengthened.
Undoubtably, the emerging dominance of behaviorism in many fields
lent Hebb's rule a certain
amount of intellectual support.
Hebb's rule is also appealing from a genetic point of view,
since it requires very little genetic ``overhead"
to implement in actual nervous systems.
All that is required is
a mechanism for distinguishing simultaneous
stimuli at the cellular level.
\skip
Verification has taken time,
but there is now evidence that Hebbian-type long term
potentiation (LTP) (with some modifications of the original
hypothesis) does indeed occur [Lynch 1986; Kennedy 1988; Stevens
1989].
Long-term depression (LTD) has been
observed in the same system supporting an ancillary
``Hebbian covariance learning rule" [Stanton and Sejnowski 1989].

\skip
\leftline{\bigbf 2.1 Experimental Evidence: The NMDA Receptor}
\skip
Long-term potentiation is mediated by the
N-methyl-D-aspartate (NMDA) receptor.
It is useful to review the mechanisms
current model of LTP for two reasons.
First, it illustrates how the proposed (Hebbian)
learning rule influenced experimental efforts.
Secondly, the actual mechanisms discovered
are subtly different from the Hebbian ideal
of strengthening correlated inputs.
\skip
According to the current model of LTP
[Zalutsky and Nicoll 1990; Buonomano and Bryne 1990;
Kandel and O'Dell 1992],
for the NMDA receptor channel to open,
two conditions must be met simultaneously:
(i) the receptor must bind glutamate, and
(ii) the postsynaptic cell must be depolarized
through activation of non-NMDA receptors.
At resting
potential, the NMDA receptor channel is blocked by Mg$^{2+}$.
Depolarization removes the voltage-dependent
Mg$^{2+}$ block, allowing
Ca$^{2+}$ to flow into the cell.
Ca$^{2+}$ appears to trigger LTP,
through the activation of at least three
different protein kinases (see Fig. 1).
\skip
There is also evidence for chemical and/or structural
{\it presynaptic} changes
[Zalutsky and Nicoll 1990; Edwards 1991].
Presynaptic modification
is thought to be
effected via retrograde messengers
released across the synaptosomal junction.
The retrograde messenger is presumed to
be a labile, diffusible substance
synthesized and released by the postsynaptic cell.
The synthesis and/or release of such messengers is likely
to be a calcium-dependent process as well.
Several substances have been postulated to function
as retrograde messengers. Among them are
nitric oxide [Gally et al. 1990],
hydrogen peroxide [Colton et al. 1989; Zoccarato et al. 1989]
and archidoinic acid [Williams et al. 1989].
(For a review, see Montague et al. [1991].)

\skip
Many other substances have been shown to
have modulatory effects on LTP.
A partial list of proteins, hormones, neurotransmitters
and other compounds includes
glycine and D-serine [Salt 1989],
serotonin [Ropert and Guy 1991],
acetylcholine and noradrenaline [Bear and Singer 1986;
Brocher et al. 1992],
human epidermal growth factor [Abe and Saito 1992],
antidepressant drugs [Birnstiel and Haas 1991],
milacemide [Quartermain et al. 1991],
opioids [Xie and Lewis 1991] and
ethanol [Iorio et al. 1992].
Thus, it is not surprising that
mental states and other factors such as ``attention",
blood flow, ``excitement", etc. can influence learning.
That so many compounds can modulate LTP indicates that
the NMDA receptor may be a much more universal tool
for synaptic modification, and not only employed in
local, Hebbian-type learning.

\skip
Finally, NMDA clearly mediates some,
but not all, forms of learning.  For instance,
Malenfant et al. [1991] showed that application of
an NMDA receptor antagonist (MK801) could block the acquisition
of a spatial maze task in a dose-dependent manner.
However, MK801 did not block the acquisition of experience-
based maternal behavior.
The same maternal experience effects {\it can}
be blocked by chemical inhibition of protein synthesis.

\skip
In summary, the NMDA receptor requires coincident events
and makes possible a type of associative learning.
Its discovery required intricate experiments
at synaptic junctions.
It is currently unclear whether synaptic change
occurs at the postsynaptic dendritic spine,
the presynaptic glutamate axon terminal,
the presynaptic depolarizing axon, the
axonal processes themselves, or a combination
of all of these structures.
Several chemical compounds have been identified
which can facilitate
or inhibit LTP.
Many compounds which modulate
LTP are common physiological
chemical compounds, proteins
or neurotransmitters
and, as such, do not necessarily originate
from either the pre- or postsynaptic neuron(s).
Thus, it is conceivable that several forms
of learning are operating in neural tissues,
and {\it these other forms of
learning can be mediated via the NMDA receptor
as well as by other, independent neural processes.}

\skip
\leftline{\bigbf 2.2 Limitations of Hebbian Learning}
\skip
Theoretically, Hebbian learning can account for some types of
biological learning.  Hebbian mechanisms
have been shown to be sufficient
to account for topographic mappings
[Kohonen 1984; Grajski and Merzenich 1990],
plasticity in cortical representation [Merzenich et al. 1987;
Montague et al. 1991]
and, when applied to ``sigma-pi" neurons,
some nonlinear pattern recognition tasks [Mel 1992].
But there is more to
the brain than conditioned reflexes and associative memories.  For
anything but special cases, Hebb's rule is insufficient as a learning
rule [Rosenblatt 1962; Rumelhart and McClelland 1986].

\skip
Since Hebbian learning requires near simultaneous or
synchronous stimuli, it is limited {\it temporally}.
In many biological
situations, instantaneous performance results are not available.
Motor control tasks, for example, are inherently sequential.
Temporal delays are also involved in
many phenomena observed in
psychophysical and electrophysiological studies of classical
conditioning, such as anticipation of an
unconditioned stimulus
[Chester 1990; Deno 1991].
Hebbian learning would have to be combined with
additional memory mechanisms
or neuronal structures
to account for such phenomena.
Recent attempts to expand Hebbian
learning rules to include short-term memory [Sutton and Barto 1983,
Klopf 1989, Grossberg and Schmajuk 1989] have met with limited
success [Chester 1990].

\skip
To account for more complex phenomenon, such as skilled movement,
many have postulated the brain utilizes ``model-reference control",
that is, the brain develops an internal model of the musculature
and environment to predict performance of a control signal.
A Hebbian mechanism can then be used to
control such a system, since presumably, the temporal delay
has been removed from correlated events.
Such a system may in fact be used, especially for
rapid, open-loop eye and hand movements
[Crossman 1983; Anderson and Vemuri 1992].
But the ``model" must still be updated by a global supervisory
signal which takes its cues from the external environment.

\skip
Since the Hebbian rule applies only to correlations at the
synaptic level, it is also limited {\it locally}.
Strengthening a local
correlation in the context of a
nonlinear mapping of several variables
(such as the N-bit parity problem) often reduces overall
performance.
Consequently, Hebbian learning is unable to reliably
train a multilayer perceptron
network to learn arbitrary, nonlinear
decision boundaries
[Rumelhart and McClelland 1986].

\skip
\leftline{\bigbf 3. THEORETICAL LEARNING RULES}
\skip
We have seen how influential
a simple theoretical concept,
Hebb's rule, has been in neurobiology.
Current artificial neural network (ANN) research
has provided valuable insights into
the collective behavior
of small networks of neurons
[Hopfield 1984; Lehky and Sejnowski 1988, 1990; Lockery et al. 1989].
However, most of these results were obtained using more
sophisticated algorithms than Hebb's rule.
Do any of the multitude of ANN learning rules
have any implications for experimental neurobiology?

\skip
Learning rules employed to train ANN's
are more appropriately referred to as optimization procedures.
These algorithms,
most of which are based on minimization of a
defined error function,
are capable of overcoming the limitations
of Hebb's rule.
Among the most popular today are genetic algorithms
[Montana and Davis 1989; Austin 1990] and
gradient-descent learning [Rumelhart et al. 1986].
(For an overview of ``connectionist" learning rules,
see Hinton [1989].)
Most of these algorithms have little biological basis
and are used primarily for engineering problems
in pattern recognition, classification,
signal reconstruction, and so on.

\skip
Criticisms of the biological plausibility of ANN
training algorithms are abundant in the literature.
In his article ``The recent
excitement about neural networks", Francis
Crick [1989a] writes:

``It is hardly surprising that such achievements [referring to the
successes using back-propagation] have produced a
heady sense of euphoria.  But is this what the brain actually
does? Alas, the back-prop nets are unrealistic in almost every
respect....{\it Obviously what is really required is a brain-like
algorithm which produces results of the same general character
as back propagation}"[emphasis added].

Bartlett Mel [1990] poses the problem this way:

``[I]s it...a fundamental law that neural associative learning
algorithms must be either representationally impoverished or
mechanistically overcomplex?"

\skip
What are the necessary features of a biologically plausible
learning rule?
First, it must have a mechanism for synaptic
modification that is consistent with experimental data.
Secondly, a learning rule must not involve so much
specific neural structure that an excessive number of genes
are required for its coding.
Lastly, to be of any use to biologists, it must be {\it observable}.
Clearly, Hebb's rule satisfies these criteria,
while back-propagation violates all three.
As the title of this paper suggests, there is at least
one other
ANN learning rule that satisfies these criteria -
a biased random-walk
[Bremermann and Anderson 1989, 1991].

\skip
\leftline{\bigbf 3.1 Learning via Random-Walks}
\skip

In its most basic form,
a random-walk can be generated by
spontaneous, random variation in synaptic strength.
This way,
the mechanism for synaptic change
is local and independent of any
higher-level teaching signals.
Successful changes in architecture or synaptic strength
are rewarded or punished {\it after the fact}.
Such a biased random-walk in synaptic weight space
can be considered a cellular analog of trial-and-error.

\skip
The first attempt to
apply such an algorithm to neural networks
was by Lewey Gilstrap, Jr., Cook and Armstrong
at Adaptronics, Inc.
(McLean, VA) around 1970
They called their algorithm ``accelerated, guided random
search" (GARS):
\skip
``[T]he accelerated random search begins by exploring the
vicinity of its initial estimate.  The random trials are governed
by a normal distribution of probabilities which is centered on
the initial point. ... the accelerated random search follows an
unsuccessful random step, with a step of equal magnitude in
the opposite direction.  By this means, a successful step is
usually achieved on the second trial if not on the first random
trial. ... A successful step is always followed by another step in
the same direction ... each successive step is given double the
magnitude of the prior step."[Barron 1968]
\skip
Barron [1968; 1970] used GARS to optimize control parameters in
flight control systems.  Mucciardi [1972] applied GARS to
neural net-like classification structures called ``neuromine nets".
Mucciardi's paper presented an analysis of neuromine nets and the
algorithm, but provided only simple examples of its application.
Interest in neural networks was waning at that time,
especially because of well-known limitations of
simple perceptrons acknowledged by Rosenblatt [1962]
and highlighted in {\it Perceptrons} [Minsky and Papert 1969].
Unfortunately, Mucciardi and his colleagues never
applied their algorithm to the complex classification problems
emphasized in {\it Perceptrons}
- the exclusive OR and ``connectedness" problems.
Another aspect of random search, overlooked by the group
at Adaptronics, was its potential relevance
to biology.

\skip
In 1988, we began experimenting with a similar
algorithm, which we dubbed the ``chemotaxis algorithm"
[Bremermann and Anderson 1989, 1991; see also Appendix],
by analogy to the strategy employed by bacteria
to find chemoattractants in a
spatial concentration gradient
[Alt 1980; Koshland 1980; Berg 1983].
We showed that a biased Gaussian random-walk could,
in fact,
train neural networks to solve the same difficult
Boolean mappings that had eluded single layer
perceptrons and Hebbian networks
(exclusive OR, N-bit parity, etc.).

\skip
Random-walk learning has not received much
attention for several reasons:

Criticism \#1: {\it Random walks were known to get trapped in local minima
in conventional optimization problems.}

In the case of neural networks, local minima
is not as much
of a problem as one might expect.
What is a local minimum in a small
network with a lower dimensional
weight space, often becomes a multi-dimensional saddle point
in higher dimensions
[Baldi and Hornik 1989; Conrad and Ebeling 1992; Yu 1992].
This is because of the degeneracy inherent in neural network
architectures:
there are usually a much larger number of free parameters (weights)
than are theoretically required to solve the task at hand.

\skip
Evolutionary optimization is also
easier in high-dimensional, redundant systems
[Conrad 1983].
A biased random-walk can be considered a rudimentary
genetic algorithm -- where the environment selects one of
two possible mutant structures at each step.
Conrad and Ebeling [1992] have shown that saddle points,
not isolated peaks, dominate high dimensional fitness landscapes:
``Increasing the dimensionality of a system...increases the chances
of finding an uphill [favorable] pathway to still higher
peaks."
Conrad refers to this phenomenon as ``extradimensional
bypass".

\skip
Criticism \#2: {\it A random walk was thought to be inefficient.}

A biased random walk is also a
form of gradient descent
(random descent), and is quite efficient.
In the case of a 3-dimensional spherical gradient
(a condition that is ideal for gradient descent),
the path taken to reach the optimum by
the chemotaxis algorithm is, on average,
only 39\% longer than the optimal, direct gradient path
[Bremermann 1974].
Empirical studies show that the
chemotaxis algorithm, while usually slower to converge,
compares favorably in final network performance
with back-propagation on a variety of benchmark tasks
[Bremermann and Anderson 1989; Wilson 1991].
Furthermore, in cases where local minima {\it do} exist,
there is no reason to expect it is more prone to local minima
than back-propagation [Anderson 1991; Baldi 1992].
An extensive analytical comparison of random
descent and gradient descent
learning is given by Baldi [1992].

\skip
Criticism \#3: {\it Neural network researchers generally
did not believe
a random walk could train neural networks
to solve complex, nonlinear
mappings such as the exclusive OR.}

The perceived
problem of local minima reinforced this belief.
This belief, however, turned out simply to be unfounded
[Bremermann and Anderson 1989] (Table I).
In addition to the benchmark problems,
the chemotaxis algorithm has been applied successfully
to training neural networks to solve
a variety of
problems: discrimination of seismic signals
[Dowla et al. 1990; Anderson 1991],
training ``recurrent" neural networks
[Anderson 1991],
process control [Willis et al. 1991a,b],
and motor control [Anderson and Vemuri 1992; Styer
and Vemuri 1992a,b].
Experiments with other
stochastic training algorithms
have had similar successes
[Harth and Tzanakou 1974;
Tzanakou et al. 1979;
Harth et al. 1988;
Smalz and Conrad 1991; Jabri and Flower 1992].

\skip
Criticism \#4: {\it ``Reinforcement" learning models had not been presented in
a
distilled, biologically plausible way.}

Reinforcement signals are generally thought
to carry only general information about
the overall performance (``good", ``better",
``target was missed by x amount", etc.).
Specific information to individual synapses
as to their relative responsibility in the
task would be very difficult to determine.
Biological mechanisms for assigning responsibility
to each individual synapse is highly unlikely [Crick 1989a].

\skip
Most proposed reinforcement learning rules
are ``mechanistically overcomplex".
In Barto and Sutton's reinforcement
learning schemes, for example,
synaptic change is
generated by the reinforcement signal itself,
as interpreted by
an adaptive critic element
[Barto et al. 1981; Barto and Sutton 1983].
Although this work has generated many interesting and non-trivial
applications, the complexity of its synaptic adjustment
rules makes it an unlikely candidate for a biological learning rule.
Other reinforcement algorithms have similar drawbacks
[Williams 1992].
Surprisingly, in a comparison between adaptive critic
and chemotaxis in controlling a cart-pole system,
chemotaxis performed as well or {\it better}
than the more complicated (and less biological)
adaptive critic networks
[Styer and Vemuri 1991a,b].

\skip
Criticism \#5: {\it Experimentalists are limited by what is observable.}

The final, and most important obstacle to finding
biological evidence for reinforcement learning has been,
and continues to be {\it experimental observability}.
This is because random walks are a non-local phenomenon.
Experimental protocols involving
single neurons, synapses, or
even a small collection of interacting neurons
cannot {\it directly} verify a non-local learning rule.
Local measurements of a global phenomenon
can only verify two of the
necessary elements, local synaptic variation
and neuromodulation (facilitation or inhibition
of synaptic change).
We devote the majority of the remainder of this article to
addressing this problem.

\skip
\leftline{\bigbf 4. BIOLOGICAL EVIDENCE}
\skip

Reinforcement learning requires three components:
(i) a mechanism for the generation of synaptic change,
(ii) a structure for evaluating performance, or ``trainer",
and
(iii) a reinforcement signal.
To build a case for biological plausibility,
we must show that all of the necessary elements
are {\it consistent} with biological observations.

\skip
Two components required for random-walk learning
are clearly consistent with biological observations -
random synaptic variation and neural structures for evaluating
performance.
Indeed, it is generally believed that local
random explorations account for some types
of neural development [Montague et al. 1991].
In developmental models, however, the reinforcement signal
is provided by the {\it target cell}.
The random walk ends when
a process finds its target.
This type of locally reinforced random-walk
has the same limitations as Hebbian learning.
The difference with what is being proposed here
is that the reinforcement signals are not
generated locally, through retrograde messengers
or cell-adhesion molecules.
Instead, reinforcement is
generated and broadcast from ``supervisory"
neural structures (Figure 2).

\skip
\leftline{\bigbf 4.1 Random Structural Variation}
\skip
Cellular events are dominated by stochastic processes.
It is highly probable that the organism makes
use of this fact in the process of learning.
It has been shown that structural variation can be guided
or influenced by chemical or neural signals.
What remains to be found is if this modulation
is a local phenomenon or mediated by higher centers.
Here, we cite just two examples of experimental systems
which are consistent with this view.

\skip
Growth of neurites in cerebellar granule cell cultures
progresses stochastically
[Rashid and Cambray-Deakin 1992].
Stimulation with NMDA results in a marked increase
in growth rate, while the addition
of an NMDA receptor antagonist,
aminophosphonovalerate (APV), causes
a marked retraction of pre-existing processes.
Either of these effects could be directed from
more distant neural structures.

\skip
In another experiment,
Glanzman et al. [1990] studied an {\it in vitro}
coculture of {\it Aplysia}
sensory neurons and their target (L7 motor)
cells. The sensorimotor cocultures were grown for 5 days and
observed by fluorescence video micrographs. One group of
preparations was repeatedly treated with the facilitating transmitter
serotonin (5-HT) for 24 hours.
At the end of the experiment, the coculture was imaged
again to look for structural changes.
Morphological
changes (changes in the size of varicosities or new processes)
at the junctions between the
sensory and motor cells were rated on a subjective scale.
This study was significant in that they were able
to {\it directly}
image structural changes - rather than relying on
comparisons between two different populations of neurons.
In the control group, morphological changes were found to be
normally distributed with a mean change of zero on their rating
scale. In the cocultures treated with serotonin, however, structural
change was shown to be highly biased toward increases in
varicosities or processes. Furthermore, they showed that these
structural changes corresponded to measurable changes in
monosynaptic excitatory post-synaptic potential (EPSP) produced in
L7 motor cells by firing the sensory neuron. Thus, they were able to
show both physical and electrophysiological facilitation can be
induced in vitro by a single chemical signal - serotonin.

\skip
We suggest that these random variations
serve a vital role in learning, that is, generating
new trial connections and efficacies.
Serotonin release in a cluster of neurons
{\it may} serve as a
local ``print" (or fixing) signal to retain effective
changes. However, the experiment described by Glanzman et al.
cannot differentiate between serotonin's putative role as a
simple growth factor or a reinforcement signal.

\skip
Serotonin has been shown to serve a role
as a neuromodulator as well as a facilitation signal.
There is evidence for a brainstem serotonergic
projection to the ventrobasal thalamus,
thus linking facilitory
signal to higher brain centers
[Eaton and Salt 1989].
Does facilitation reinforce existing changes,
or does the change occur as a result of the
presence of serotonin?

\skip
\leftline{\bigbf 4.2 Reinforcement Signals}

\skip
A biased random-walk requires that the performance of a net be
evaluated. This evaluation could be accomplished by other brain
circuits.  We do not consider this requirement problematic, since
evaluation of performance tends to be computationally easier than
improvement. For example, throwing a ball requires precise
coordination and timing of numerous muscles. Good performance is
hard to achieve and may require extensive training.  But, how close a
ball comes to hitting the target is relatively easy to determine.
Evaluation of accuracy can be processed separately by the visual
cortex - independent of networks involved in generating the
movement.  One portion of the brain thus could act for another
system as ``supervisor".

\skip
The reinforcement signal is likely to
carry only general, non-specific, information.
Thus, it could be neural or
chemical (hormonal) in origin.
Many of the substances which have been shown to modulate
LTP (including the candidate retrograde messengers) are
candidate reinforcement signals as well.
To complete a model of random-walk learning,
one must demonstrate that other brain centers
have projections to the sites of synaptic variation
which release (directly or indirectly)
substances which can act to facilitate or
inhibit the process of structural change.

\skip
One known reverse pathway is a projection from
the locus coeruleus to the olfactory bulb.
Locus coeruleus neurons are known to have
norepinephrine (NE) as a neurotransmitter.
Gray et al. [1986] demonstrated that intrabulbar
infusion of NE into the rabbit olfactory bulb can
prevent or delay the habituation to unreinforced odors.
Locus coeruleus neurons are known
to be activated by unconditioned stimuli
[Aghajanian and Vandermaelen 1982], and
several forms of use-dependent synaptic plasticity in
cortical tissues require the presence of NE
[Bliss et al. 1983; Bear and Singer 1986].
These signals from the locus coeruleus are diffuse
but may still serve a neuromodulatory role [Crick 1989a].
Taken together, these data suggest that norephinephrine
could be functioning as a reinforcement signal.

\skip
\leftline{\bigbf 5. CONCLUSIONS}

\skip
It is self-evident that
some form of trial-and-error learning is involved
in the acquisition of skilled movement
[Crossman 1959; Anderson 1981].
But training a {\it tabula rasa} of randomly connected
masses of neurons
to perform complex control tasks is evidently a hopeless
endeavour [Anderson 1991].
High level control of movement is thought to involve
the coordination or modulation of existing Central
Pattern Generators (CPG's) [Selverston 1980].
A biased random-walk can be used
to optimize crudely organized network of
CPG's during the acquisition of skilled movement
[Anderson 1991; Anderson and Vemuri 1992; Styer and Vemuri 1992a,b].
This is somewhat analogous to Edelman's selectionist
hypothesis in that learning entails the ``selection", or education
of an existing repertoire of dynamical ``groups"
[Edelman 1987; Crick 1989b].
Furthermore, we point out that the chemotaxis algorithm
is the most primitive form of trial-and-error; undoubtedly,
more sophisticated, higher level neural mechanisms
will
have evolved to coordinate and compliment this
process [Smalz and Conrad 1991].

\skip
Experimental verification of this type of learning will
require protocols involving {\it collections or assemblies}
of neurons, rather than individual synaptic junctions,
to observe the stochastic variation and the effects
of putative reinforcement signals.
Furthermore, a more ambitious effort must be
made to link reinforcement signals backwards
to their {\it projective} sources.

\skip
McCulloch and Pitts offered a solution to the embodiment problem
by demostrating the computational properties of neural networks.
Hebb proposed an neurbiological correlate to
associative learning or classical conditioning.
Biased random-walks in synaptic weight space
can be seen as the neurobiological ``embodiment"
of trial-and-error learning.
A biased random walk may some day be shown to be
``{\it a learning rule immanent in nervous activity}".

\skip
\leftline{\bigbf Acknowledgements}
\skip
I thank Daniel Chester for calling to my attention
the work done at Adaptronics, Inc..
I also thank Hans J. Bremermann, Lee Segel
Michael Conrad and V. (Rao) Vemuri for
their encouragement and editorial comments.
This work was performed under the auspices of the U.S.
Department of Energy and supported by
the Center for Nonlinear Studies at Los Alamos.

\vfill\eject
\skip
\centerline{\bf APPENDIX}
\skip

\skip
\leftline{\bf The Chemotaxis Algorithm}

\skip
The ``chemotaxis training algorithm" consists of a biased
random-walk in weight space. One advantage to this training
method is that it does not require gradient calculations or detailed
error signals. It also allows for automatic adjustment of the
single learning parameter, which otherwise has to be found
empirically.

\skip
The network is initialized with an an arbitrary set of weights,
$w^o$, and performance E($w^o$) is evaluated.
A random vector $\Delta w$ is
chosen from a multivariate Gaussian distribution with a zero mean
and a unit standard deviation.  This random vector is added to the
current weights to create a ``tentative" set of weights (w$^t$):
$$ w^t = w^o + h \Delta w $$
where h is a stepsize parameter. Performance E($w^t$) is then
calculated for the tentative weights.  If the error of the new
configuration is lower than the original configuration,
the tentative
changes in the weight vector are retained; otherwise, the system
reverts to its original configuration.  If a successful direction in
weight space is found, weight modifications continue along the
same random vector until progress ceases.
A new random vector is
then chosen, and the process is repeated.
More details are  available in the cited
literature.

\vfill\eject
\skip
\centerline{\bf TABLE AND FIGURE CAPTIONS}
\skip

\skip
\leftline{\bigbf Table I: Training Time for the N-bit Parity Problem.}
\skip
N-bit parity can be considered a generalization of the 2-bit
``exclusive OR" (XOR) problem since class membership of a given
pattern is dependent on all N inputs.
Network architecture was N-(2N+1)-1, where N represents
the number of hidden units. The networks were trained on
all $2^N$ possible binary input patterns. Training was
continued until the network responses were within
10\% of the ideal Boolean values.
Chemotaxis averages are taken from Bremermann and Anderson [1989].
No attempt was made to optimize algorithm parameters.
Backpropagation averages are taken from Tesauro and Janssens [1988],
who used optimal values for the learning and momentum parameters.
Note that the {\it computational effort} is double these values in the case of
backpropagation.

\skip
\leftline{\bigbf Figure 1: NMDA Implementation of Hebbian Learning.}
\skip
Simultaneous membrane depolarization and activation
of the NMDA receptor allows calcium ions to flow into
the cell.
Calcium dependent proteins trigger a cascade of
intracellular events leading to structural and/or
chemical changes postsynaptically as well
as potential presynaptic changes via retrograde
messengers.
(Adapted from [Montague et al. 1991; Kandel and O'Dell 1992].)

\skip
\skip
\skip
\leftline{\bigbf Figure 2: Neural Implementation of a Biased Random-Walk}
\skip
Random variation in synaptic connectivity and efficacy
is rewarded {\it after the fact} if performance has improved.
Performance is evaluated by sensory systems (somatosensory,
visual, auditory, etc.) and a non-specific, reinforcement
signal is broadcast to the participating neural circuitry.
The reinforcement signal could be chemical (hormonal)
or neural in origin.
\vfill\eject
\skip

\hsize=6.5truein
\vsize=8.50truein
\raggedbottom

\tolerance=10000
\parindent=0pt
\clubpenalty=10000
\widowpenalty=10000

\hbox { }
\baselineskip=12pt
\centerline{\bf Table I}
\bigskip
\centerline{\bf Chemotaxis Algorithm Performance}
$$\vbox{\tabskip=0pt
\halign  {
\hfill#\tabskip.5truein&
\hfill#\tabskip.15truein&
\hfill#\tabskip.15truein \tabskip0pt\cr
\noalign{\vskip.15truein}
\noalign{\vskip.15truein}
\noalign{\hrule}
\noalign{\vskip.15truein}
Dimension & Chemotaxis & Backpropagation \cr
(N) & (epochs) & (epochs)\cr
\noalign{\vskip.15truein}
\noalign{\hrule}
\noalign{\vskip.15truein}
2(XOR) & 113 & 25 \cr
3 & 251 & 33 \cr
4 & 962 & 75 \cr
5 & 1259 & 130 \cr
6 & 4169 & 310 \cr
7 & 5789 & 800 \cr
\noalign{\vskip.15truein}
\noalign{\hrule}
\noalign{\vskip.15truein}
& \cr
}}$$

\vfill\eject
\centerline{\bf LITERATURE}
\bigskip
{
\frenchspacing
\parskip=8pt
\def\ti#1{#1.} 
\def\bt#1{{\sl #1}} 
\def\vo#1{{\bf #1}} 
\def\jo#1{{\sl #1},} 

K. Abe and H. Saito (1992).
\ti{Epidermal growth factor selectively enhances NMDA receptor-mediated
increase of intracellular Ca2+ concentration in rat hippocampal neurons}
\jo{Brain Research}
\vo{587}:102-8.

G. K. Aghajanian and C. P. Vandermaelen (1982).
\ti{Intracellular identification of central noradrenergic and
serotonergic neurons by a new double labeling procedure}
\jo{J. of Neuroscience}
\vo{2}:1786-1792.

W. Alt (1980).
\ti{Biased random walk models for chemotaxis and
related diffusion approximations}
\jo{J. Mathem. Biology}
\vo{9}:147-177.

J. R. Anderson, Ed. (1981).
\ti{Cognitive Skills and Their Acquisition}
Lawrence
Erlbaum Associates, Hillsdale, N.J.

R. W. Anderson and V. Vemuri (1992).
\ti{Neural Networks can be used
for Open-Loop, Dynamic Control}
\jo{Int. J. Neural Networks}
\vo{2}(3)
(Abstract in: Proc. Int. AMSE Conf. Neural Networks, San
Diego, CA,
\vo{2} pp. 227-237 (May 29-31, 1991).

R. W. Anderson (1991).
\bt{Stochastic Optimization of Neural Networks and Implications
for Biological Learning}
Ph.D. Dissertation, University of California, San Francisco.

M. A. Arbib (1987).
\bt{ Brains, Machines, and Mathematics, Second Edition}
(First Edition: McGraw 1964), Springer-Verlag, New York.

S. Austin (1990).
\ti{Genetic Solutions To XOR Problems}
\jo{AI Expert} pp. 52--57

P. Baldi and K. Hornik (1989).
\ti{Neural Networks and Principle Component Analysis:
Learning from Examples Without Local Minima}
\jo{Neural Networks}
\vo{2}:53-58.

P. Baldi (1991).
\ti{Gradient Descent Learning Algorithms:
A General Overview}
JPL Technical Document.

R. L. Barron (1968).
\ti{Self-Organizing and Learning Control Systems}
in:
Cybernetic Problems in Bionics (Bionics Symposium, May 2-5, 1966,
Dayton, Ohio), New York, Gordon and Breach, pp. 147-203.

R. L. Barron (1970).
\ti{Adaptive Flight Control Systems}
In: Principles and
Practice of Bionics (NATO AGARD Bionics Symposium, Brussels,
Belgium, Sept. 18-20, 1968), pp. 119-167.

A. G Barto, R. S Sutton and P. S Brouwer (1981).
\ti{Associative
Search Network: A Reinforcement Learning Associative Memory}
\jo{Biological Cybernetics}
\vo{40}:201-211.

A. G Barto and R. S. Sutton (1983).
\ti{Neuronlike Adaptive
Elements That Can Solve Difficult Learning Control Problems}
\jo{IEEE
Transactions on Systems, Man, and Cybernetics}
\vo{SMC-13} (5):835-846.

M. F. Bear and W. Singer (1986).
\ti{Modulation of visual cortical plasticity by acetylcholine
and noradrenaline}
\jo{Nature}
\vo{320}:172-17.

H. Berg (1983).
\bt{Random Walks in Biology}
Princeton University Press,
Princeton.

S. Birnstiel and H. L. Haas (1991).
\ti{Acute effects of antidepressant drugs on long-term
potentiation (LTP) in rat hippocampal slices}
\jo{Naunyn-Schmiedebergs Archives of Pharmacology}
\vo{344}:79-83.

W.W. Bledsoe (1961a).
|ti{The Use of Biological Concepts in the Analytical Study
of Systems}, Technical Report, Panoramic Research Inc., Palo Alto, CA

T. V. Bliss, G. V. Goddard and M. Riives (1983).
\ti{Reduction of long-term potentiation in the dentate gyrus
of the rat following selective depletion of monoamines}
\jo{J. of Physiol.}
\vo{334}:475-491.

H. J. Bremermann (1974).
\ti{Chemotaxis and Optimization}
\jo{ J. of the Franklin Institute}
(Special Issue: Mathematical Models of Biological Systems)
\vo{297}:397-404.

H. J. Bremermann and R. W. Anderson (1989).
\ti{An Alternative to
Back-propagation: A Simple Rule of Synaptic Modification For Neural
Net Training and Memory}
Technical Report: U. C.
Berkeley Center for
Pure and Applied Mathematics PAM-483.

H. J. Bremermann and R.	W. Anderson (1991).
\ti{How the Brain Adjusts Synapses - Maybe}
In:Automated Reasoning: Essays in Honor
of Woody Bledsoe, R. S. Boyer (ed.), Chapter 6, pp. 119-147, Kluwer
Academic Pub., Boston.

S. Brocher, A. Artola and W. Singer (1992).
\ti{Agonists of cholinergic and noradrenergic
receptors facilitate synergistically the induction of
long-term potentiation in slices of rat visual cortex}
\jo{Brain Research}
\vo{573}:27-36.

D. V. Buonomano, and J. H. Bryne (1990).
\ti{Long-Term Synaptic Changes
Produced by a Cellular Analog of Classical Conditioning in Aplysia}
\jo{Science}
\vo{249}:420-3.

D. L. Chester (1990)
\ti{A Comparison of some Neural Network Models of
Classical Conditioning}
\jo{Proc. 5th IEEE Int. Symposium on Intelligent
Control} Philadelphia, PA,
\vo{2}:1163-1168.

C. A. Colton, L. Fagni and D. Gilbert (1989).
\ti{The action of hydrogen peroxide on paired pulse and
long-term potentiation in the hippocampus}
\jo{Free Radical Biol. Med.}
\vo{7}:3-8.

M. Conrad (1983).
\bt{Adaptability} (Chapter 10), Plenum Press, N.Y.

M. Conrad and W. Ebeling (1983).
\ti{M.V. Volkenstein, evolutionary thinking and the structure of
fitness landscapes}
\jo{Biosystems}
\vo{27}:125-128.

F. Crick (1989a).
\jo{The Recent Excitement about Neural Networks}
\jo{Nature}
\vo{337}:129-132.

F. Crick (1989b).
\ti{Neural Edelmanism}
\jo{Trends in Neurosciences}
\vo{12}
(7):240-248.

E. R. F. W. Crossman (1959).
\ti{A Theory of the Acquisition of Speed-Skill}
\jo{Ergonomics}
\vo{2}(2):153-166.

E. R. F. W. Crossman and P. J. Goodeye (1983).
\ti{Feedback Control of Hand-
Movement and Fitt's Law}
\jo{Quarterly Journal of Experimental
Psychology}
\vo{35A}:251-278.

Y. Dan and M. Poo (1992).
\ti{Hebbian Depression of Isolated Neuromuscular
Synapses in Vitro}
\jo{Science}
\vo{256}:1570-1573.

K. Deno (1991).
Ph.D. Thesis, Dept. EECS, U.C. Berkeley.

F. U. Dowla, S. R. Taylor and R. W. Anderson (1990).
\ti{Seismic
Discrimination with Artificial Neural Networks: Preliminary Results
with Regional Spectral Data}
\jo{Bull. Seismological Society of
America}
\vo{80}(5):1346-1373.

S. A. Eaton and T. E. Salt (1989).
\ti{Modulatory Effects of Serotonin on
Excitatory Amino Acid Responses and Sensory Synaptic Transmission
in the Ventrobasal Thalamus}
\jo{Neuroscience}
\vo{33}(2,):285-292.

G. M. Edelman  (1987).
\bt{Neural Darwinism} Basic Books, New York.

F. Edwards (1991).
\ti{LTP is a long term problem}
\jo{Nature}
\vo{350}:271-272.

J. A. Gally, P. R. Montague, G. N. Recke and G. M. Edelman
(1990).
\ti{The NO hypothesis: possible effects of a rapidly diffusible
substance in neural development and function}
\jo{Proc. Natl. Acad. Sci. USA}
\vo{87}:3547-3551.

D. L. Glanzman, E. R. Kandel and S. Schacher (1990).
\ti{Target-
Dependent Structural Changes Accompanying Long-Term Synaptic
Facilitation in Aplysia Neurons}
\jo{Science}
\vo{249}:799-802.

K. A. Grajski and M. M. Merzenich (1990).
\ti{Hebb-Type Dynamics is
Sufficient to Account for the Inverse Magnification Rule in Cortical
Somatotopy}
\jo{Neural Computation}.

C. M. Gray, W. J. Freeman and J. E. Skinner
(1986).
\ti{Chemical Dependencies of Learning in the Rabbit Olfactory
Bulb: Acquisition of the Transient Spatial Pattern
Change Depends on Norepinephrine}
\jo{Behavioral Neuroscience}
\vo{100}(4):585-596.

S. Grossberg and N. A. Schmajuk (1989)
\ti{Neural dynamics of
adaptive timing and temporal discrimination during associative
learning}
\jo{Neural Networks} \vo{2}(2):79-102.

E. Harth, T Kalogeropoulos, A. S. Pandya and K. P. Unnikrishnan
(1988).
\ti{A Universal Optimization Network}
AT\&T Technical Memorandum \#11118-881026-23TM.

E. Harth and E. Tzanakou (1974).
\ti{A Stochastic Method for Determining Visual Receptive Fields}
\jo{Vision Res.}
\vo{14}: 1475-1482.

G. E. Hinton (1989).
\ti{Connectionist Learning Procedures}
\jo{Artificial
Intelligence} \vo{40}(1):143-150.

D. O. Hebb (1949).
\bt{The Organization of Behavior}
Wiley, New York.

J. J. Hopfield (1984).
\ti{Neurons with graded response have collective
computational properties like those of two-state neurons}
\jo{PNAS (USA)} \vo{81}: 3088-92.

K. R. Iorio, L. Reinlib, B. Tabakoff and P. L. Hoffman (1992).
\ti{Chronic exposure of cerebellar granule cells to ethanol results in
increased N-methyl-D-aspartate receptor function}
\jo{Molecular Pharmacology}
\vo{41}:1142-8.

Y. Izumi, D. B. Clifford and C. F. Zorumski (1992).
\ti{Inhibition of Long-Term Potentiation by NMDA-Mediated
Nitric Oxide Release}
\jo{Science} \vo{257}:1273-1276.

M. Jabri and B. Flower (1992).
\ti{Weight Perturbation:
An Optimal Architecture and Learning Technique for
Analog VLSI Feedforward and Recurrent Multilayer Networks}
\jo{IEEE Trans. Neural Networks} \vo{3}(1):154-157.

E. R. Kandel and T. J. O'Dell (1992).
\ti{Are Adult Learning
Mechanisms Also Used for Development?}
\jo{Science} \vo{258}:243-245.

J. A. Kauer, R. C. Malenka and R. A. Nicoll
(1988).
\ti{NMDA application potentiates synaptic
transmission in the hippocampus}
\jo{Nature}
\vo{334}: 249-252

M. B. Kennedy (1988).
\ti{Synaptic Memory Molecules}
\jo{Nature}
\vo{335}:770-772.

A. H. Klopf (1989).
\ti{Classical conditioning phenomena predicted by a
drive-reinforcement model of neuronal function}
In: Neural Models
of Plasticity: Experimental and Theoretical Approaches, J. H. Byrne
and W. O. Berry (eds.), Chapter 7, pp. 104-132, Academic Press

T. Kohonen (1984).
\bt{Self-Organization and Associative Memories} Springer-
Verlag, Berlin.

D. Koshland (1980).
\bt{Bacterial chemotaxis as a model behavioral system}
Raven Press, New York.

S. R. Lehky and T. J. Sejnowski (1988).
\ti{Computing 3-D Curvatures
from Images of Surfaces Using a Neural Model}
\jo{Nature}
\vo{333}:452.

S. R. Lehky and T. J. Sejnowski (1990).
\ti{Neuronal Model of
Stereoacuity and Depth Interpolation Based on a Distributed
Representation of Stereo Disparity}
\jo{J. of Neuroscience}
\vo{10}(7):2281-2299.

S. R. Lockery, G. Wittenberg, W. B. Kristan and G. W. Cottrell (1989).
\ti{Function of Identified Interneurons in the Leech Elucidated Using
Neural Networks Trained by Back-Propagation}
\jo{Nature}
\vo{340}:468-71.

G. Lynch (1986).
\bt{Synapses, Circuits, and the Beginnings of Memory}
Bradford/MIT Press, Cambridge, MA.

S. A. Malenfant, S. O'Hearn and A. S. Fleming,
(1991).
\ti{MK801, an NMDA antagonist, blocks acquisition of a
spatial task but does not block maternal experience effects}
\jo{Physiology and Behavior}
\vo{49}:1129-37.

W. S. McCulloch and W. Pitts (1949).
\ti{A Logical Calculus of the Ideas
Immanent in Nervous Activity}
\jo{Bulletin of Mathematical Biophysics}
\vo{5}:115-133.

B. W. Mel (1990).
\bt{Connectionist Robot Motion Planning} Academic
Press, Boston, San Diego.

B. W. Mel (1992).
\ti{NMDA-Based Pattern Discrimination in a Modeled Cortical Neuron}
\jo{Neural Computation}
\vo{4}:502-517.

M. M. Merzenich, R. J. Nelson, J. H. Kaas, M. P. Stryker, W.
M. Jenkins, J. M.
Zook, M. S. Cynader and A. Schoppman (1987).
\ti{Variability in Hand Surface
Representations in Areas 3b and 1 in Adult Owl and Squirrel
Monkeys}
\jo{J. of Comparative Neurology}
\vo{258}(2):281-96.

M. Minsky and S. Papert (1969).
\bt{Perceptrons: An Introduction to
Computational Geometry} MIT Press, Cambridge, Mass.

P. R. Montague, J. A. Gally and G. M. Edelman (1991).
\ti{Spatial Signaling in the Development and Function of
Neural Connections}
\jo{Cerebral Cortex}
\vo{1}(1):1047-3211.

D. Montana and L. Davis (1989).
\ti{Training Feedforward Neural Networks
Using Genetic Algorithms}
\jo{Proc. 11th IJCAI}.

P. G. Montarolo, E. R. Kandel and S. Schacher
(1988).
\ti{Long-Term Heterosynaptic Inhibition in Aplysia}
\jo{Nature}
\vo{333}:171-4.

A. N. Mucciardi (1972).
\ti{Neuromine Nets as the Basis for the
Predictive Component of Robot Brains} in: Cybernetics, Artificial
Intelligence, and Ecology, H. W. Robinson and D. E. Knight
(eds.), (Fourth Annual Symposium Amer. Soc. of Cybernetics),
Spartan Books, pp. 159-193.

J. C. Pearson, L. H. Finkel and G. M. Edelman (1987).
\ti{Plasticity in
the Organization of Adult Cerebral Cortical Maps: A Computer
Simulation Based on Neuronal Group Selection}
\jo{J. of Neuroscience}
\vo{7}(12):4209-4223.

N. A. Rashid and M. A. Cambray-Deakin (1992).
\ti{N-methyl-D-aspartate effects on the growth, morphology
and cytoskeleton of individual neurons in vitro}
\jo{Brain Research}
\vo{67}:301-308.

D. Quartermain, T. Nuygen, J. Sheu and R. L. Herting (1991).
\ti{Milacemide enhances memory storage and
alleviates spontaneous forgetting in mice}
\jo{Pharmacology, Biochemistry and Behavior}
\vo{39}:31-5.

N. Ropert and N. Guy (1991).
\ti{Serotonin facilitates GABAergic transmission
in the CA1 region of rat hippocampus in vitro}
\jo{Journal of Physiology}
\vo{441}:121-36.

F. Rosenblatt (1958).
\ti{The Perceptron, a probabilistic model for
information storage and organization in the brain}
\jo{Psych. Rev.}
\vo{62}:386-408.

F. Rosenblatt (1962).
{\ti Principles of Neurodynamics}
Spartan Books,
Washington, D. C..

D. E. Rumelhart, G. E. Hinton and R.J. Williams (1986).
\ti{Learning
Internal Representations by Error Propagation} In: Parallel
Distributed Processing Vol.1,
D. E. Rumelhart and J. L. McClelland, eds.,
MIT Press, Cambridge, MA pp. 318-362.

T. E. Salt (1989).
\ti{Modulation of NMDA receptor-mediated responses
by glycine and D-serine in the rat thalamus in vivo}
\jo{Brain Research}
\vo{481}:403-6.

A. I. Selverston (1980).
\ti{Are Central Pattern Generators
Understandable?}
\jo{Behavioral and Brain Sciences}
\vo{3}: pp. 535-571.

H. Seigelman and E. Sontag (1991).
\ti{Neural Nets are Universal
Computing Devices}, Technical Report SYCON-91-08, Rugters
University, Center for Systems and Control.

R. Smalz and M. Conrad (1991).
\ti{A Credit Apportionment
Algorithm for Evolutionary Learning with Neural Networks}
In: Neurocomputers and Attention Vol.II: Connectionism
and Neurocomputers,
A. V. Holden and V. I. Kryukov, eds.,
Manchester University Press: New York,
pp. 663-673.

P. K. Stanton and T. J. Sejnowski (1989).
\ti{Associative long-term
depression in the hippocampus induced by hebbian covariance}
\jo{Nature}
\vo{339}:215-218 (1989).

C. F. Stevens (1989)
\ti{Strengthening the synapses}
\jo{Nature}
\vo{338}:460-461.

D. L. Styer and V. Vemuri (1992a).
\ti{Adaptive Critic and Chemotaxis in Adaptive Control}
Conf. Artificial Neural Networks in Engineering (ANNIE),
St. Louis, MO. (Nov.).

D. L. Styer and V. Vemuri (1992b).
\ti{Control by Artificial Neural Networks Using
Model-Less Reinforcement Learning}
Preprint: Biomedical Engineering Graduate Group,
University of California, Davis, CA, USA 95616.

R. S. Sutton and A. G. Barto (1981).
\ti{Toward a Modern Theory of
Adaptive Networks: Expectation and Prediction}
\jo{Psychological Review}
\vo{88}(2):135-170.

G. Tesauro and B. Janssens (1988).
\ti{Scaling Relationships in Backpropagation Learning}
\jo{Complex Systems}
\vo{2}:39-44.

E. Tzanakou, R. Michalak and E. Harth (1979).
\ti{The Alopex Process: Visual Receptive Fields
by Response Feedback}
\jo{Biol. Cybern.}
\vo{35}:161-174.

J. H. Williams, M. L. Errington, M. A. Lynch and
T. V. P. Bliss (1989).
\ti{Arachidonic
Acid Induces a Long-Term Activity-Dependent Enhancement of
Synaptic Transmission in the Hippocampus}
\jo{Nature}
\vo{341}:739-42.

R. J. Williams (1992).
\ti{Simple Statistical Gradient-Following Algorithms
for Connectionist Reinforcement Learning}
\jo{Machine Learning}
\vo{8}:229-256.

M. J. Willis, C. Di Massimo, G. A. Montague, M. T. Tham and
A. J. Morris (1991a)
\ti{Artificial Neural Networks in Process engineering}
\jo{IEE Proceedings-D}
\vo{138} pp. 256-266.

M. J. Willis, G. A. Montague, C. Di Massimo,
M. T. Tham and A. J. Morris (1991b).
\ti{Non-Linear Predictive Control Using Optimization Techniques}
Proc. ACC, Boston, pp. 2788-2793.

J. M. Wilson (1991).
\ti{Back-Propagation Neural Networks: A Comparison
of Selected Algorithms and Methods of Improving Performance}
Proc. 2nd Annual Workshop Neural Networks WNN-AIND, Auburn,
Alabama (Feb. 11-13, 1991).

C. W. Xie and D. V. Lewis (1991).
\ti{Opioid-mediated facilitation of long-term
potentiation at the lateral
perforant path-dentate granule cell synapse}
\jo{Journal of Pharmacology and Experimental Therapeutics}
\vo{256}:289-96.

X. H. Yu (1992).
\ti{Can Backpropagation Error Surface Not Have Local Minima}
\jo{IEEE Trans. Neural Networks}
\vo{3}:1019-1021.

R. A. Zalutsky and R. A. Nicoll (1990).
\ti{Comparison of Two Forms of Long-Term Potentiation
in Single Hippocampal Neurons}
\jo{Science}
\vo{248}:1619-1624.

F. Zoccarato, R. Deana, L. Cavallini and A. Alexandre (1989).
\ti{Generation of hydrogen peroxide by cerebral cortex synaptosomes}
\jo{Eur. J. Biochem}
\vo{180}:473-478.

\bye